\documentclass[%
 aip,
rsi,%
 amsmath,amssymb,
 reprint,%
]{revtex4-1}

\usepackage{graphicx}
\usepackage{dcolumn}
\usepackage{bm}
\usepackage{epstopdf}
\usepackage{color}

\begin{document}

\preprint{AIP/123-QED}

\title{Brownian dynamics simulations of oblate and prolate colloidal particles in nematic liquid crystals}

\author{Neftal\'{\i} Morillo}
\email{jnmorgar@upo.es}
\affiliation{Department of Physical, Chemical and Natural Systems, Pablo de Olavide University, 41013 Sevilla, Spain}
\author{Alessandro Patti}
\affiliation{School of Chemical Engineering and Analytical Science, The University of Manchester, Manchester, M13 9PL, UK}
\author{Alejandro Cuetos} 
\affiliation{Department of Physical, Chemical and Natural Systems, Pablo de Olavide University, 41013 Sevilla, Spain}


\date{\today}

\begin{abstract}

It is well known that understanding the transport properties of liquid crystals (LCs) is crucial to optimise their performance in a number of technological applications. In this work, we analyse the effect of shape anisotropy on the diffusion of rod-like and disk-like particles by Brownian dynamics simulations. To this end, we compare the dynamics of prolate and oblate nematic LCs incorporating particles with the same infinite-dilution translational or rotational diffusion coefficients. Under these conditions, which are benchmarked against the standard case of identical aspect ratios, we observe that prolate particles display faster dynamics than oblate particles at short and long timescales. Nevertheless, when compared at identical infinite-dilution translational diffusion coefficients, oblate particles are faster than their prolate counterparts at short-to-intermediate timescales, which extend over almost three time decades. Both oblate and prolate particles exhibit an anisotropic diffusion with respect to the orientation of the nematic director. More specifically, prolate particles show a fast diffusion in the direction parallel to the nematic director, while their diffusion in the direction perpendicular to it is slower. By contrast, the diffusion of oblate particles is faster in the plane perpendicular to the nematic director. Finally, in the light of our recent study on the long-time Gaussian and Fickian diffusion in nematic LCs, we map the decay of the autocorrelation functions and their fluctuations over the timescales of our simulations to ponder the existence of mobile clusters of particles and the occurrence of collective motion.

\end{abstract}

\maketitle

\section{\label{sec:intro}Introduction}

In 1991, in his Nobel Lecture, Pierre-Giles de Gennes proposed to classify a number of physical systems under the general umbrella of Soft Matter \cite{DEG91}. These systems, which include polymers, colloids, and liquid crystals are nowadays pillars in the development of technological applications, whose full exploitation dramatically relies on the understanding of the fundamental laws driving their behavior. In particular, the very special properties of LCs, almost at the interface between those of liquids and ordered solids, are generally observed in a wide spectrum of fluids incorporating molecules (mesogens) with an anisotropic shape, such as rods and disks. In general terms, LCs flow like liquids, but they also exhibit a significant degree of orientational and/or positional order, resembling the internal arrangement of a crystal. Their thermodynamic properties can be investigated by relatively simple theories, which indicate that the main driving force for their formation has an entropic origin \cite{ONS49, DEG93, HANBOOK}.

The motivation behind the significant attention devoted to LCs pivots around their vast and diverse technological applications, spanning displays, photovoltaic solar cells, organic light-emitting diodes, field-effect transistors, thermometers, lasers, and nanowires \cite{SCH01, BUS02, OHT03, KUM06, KELLY}. Many of these applications rely on the particular structural properties of the LC mesophases. For instance, the functioning of LCDs is based on the anisotropy of optical properties in the nematic phase, while for the nanowires a key aspect is the overlap of $\pi$ orbitals due to the stacking of discotic molecules in columns. Dynamical properties are equally important and contribute to determine the performance of a material. Laschat \textit{et al.} showed that discotic mesogens are not useful as switching units in LCDs due to the fact that their LCs typically have larger viscosity than that measured in LCs of rod-like mesogens \cite{LAS07}. Although all these applications make use of molecular LCs, namely LCs whose mesogens are molecules, colloidal LCs, which by contrast consist of colloidal particles in suspension, are of key relevance to understand a number of processes at the molecular scales that are usually too fast to be studied by conventional microscopy. In other words, colloidal LCs are excellent model systems to unveil the behavior of molecular LCs. 

Molecular simulation has been crucial to understand and characterize liquid crystalline materials \cite{CAR05, WIL05}. Phase diagrams and equilibrium structures obtained in systems of prolate \cite{MCG96,BOL97} and oblate \cite{EPP84,MAR11} purely repulsive particles have been well characterized and a number of potentials, such as the Gay-Berne \cite{GAY81}, Kihara \cite{CUE03} or the more recent Gay-Berne-Kihara potential \cite{MAR05,MAR09B}, have been extensively studied for both particle geometries. Consequently, we now have a clear picture of the behavior of prolate and oblate particles and its dependence on particle aspect ratio, particle-particle interactions and concentration. Likewise, dynamical properties of LC phases have been investigated by both theory and computer simulation. Very early theoretical studies predicted an anisotropy in the diffusion of the particles in nematic fluids \cite{CHU75, HES91}, later corroborated by simulations \cite{DEM92, JAB12}. Specifically, it was concluded that, in the nematic phase of prolate particles the diffusion is faster in the direction parallel to the nematic director. In particular, the long-time diffusion coefficient parallel to the nematic director, $D_{\parallel}$, is larger than the diffusion coefficient perpendicular to the nematic director, $D_{\perp}$. The opposite tendency is observed in nematic phases of oblate particles. Nevertheless, in positionally order smectic and columnar LCs, the layer-to-layer and column-to-column diffusion results to be significantly reduced and becomes slower than the in-layer or in-column diffusion \cite{BIE08, PAT09, PAT10, MATENA, PAT11, PIE15}.

Although the thermodynamics of anisotropic particles have been extensively investigated in the past, studies on their diffusion in LCs have received significantly less attention, especially for oblate particles. The diffusion of colloidal particles is promoted by the random collision with the solvent molecules, which, according to the Einstein-Smoluchowsky theory, is controlled by the temperature and solvent viscosity combined in the infinite-dilution diffusion coefficients \cite{BER93}. These diffusion coefficients, which refer to an isolated particle in a medium, are the same as the short-time diffusion coefficients of a particle in a relatively denser suspension, before any collisions with its neighbors is produced. At longer timescales, the transport properties are mediated by the inter-particle collisions and the diffusion coefficients change. Due to the impact of particle anisotropy on the transport properties and on the consequent design of devices for the above-mentioned applications, in this work we present a computer simulation study on the diffusion of oblate and prolate particles in nematic LCs. Our aim is to compare the relative ability of prolate and oblate particles of diffusing in nematic LC phases. To this end, we impose the same values of the infinite-dilution translational or rotational diffusion coefficients for both prolate and oblate particles. This choice represents an important change with respect to past studies that assumed the same particle volume or aspect ratio \cite{CHU75, HES91}. 

The present paper is organised as follows. In Section II, we describe the model and simulation methodology employed, where the essential features of Brownian dynamics simulation are discussed. In Section III, we analyse the results for  oblate and prolate particles that have the same aspect-ratio, or the same infinite-dilution translational diffusion coefficient, or the sane infinite-dilution rotational diffusion coefficient. Finally, we present our conclusions.


\section{Model and Simulation Methods}

We have modelled both prolate and oblate particles as spherocylindrical particles. A spherocylinder is a solid of revolution obtained rotating a rectangle of elongation $L$ and width $\sigma$ capped by two semicircles of diameter $\sigma$ at both ends (see Fig.\ 1.a). If this 2D shape rotates around its longitudinal axis of symmetry, the prolate spherocylinder is obtained, whereas if the rotation is perpendicular to this axis, the three-dimensional body generated is an oblate spherocylinder. According to these considerations, the shape anisotropy or aspect ratio for prolate particles is defined as $a_p= (L+\sigma)/\sigma$, while for oblates reads $a_o=\sigma/(L+\sigma)$. In Fig.\ 2.b several oblate and prolate geometries are shown along with the case of a sphere. 

\begin{figure}
\center
	\includegraphics[width=0.48\textwidth]{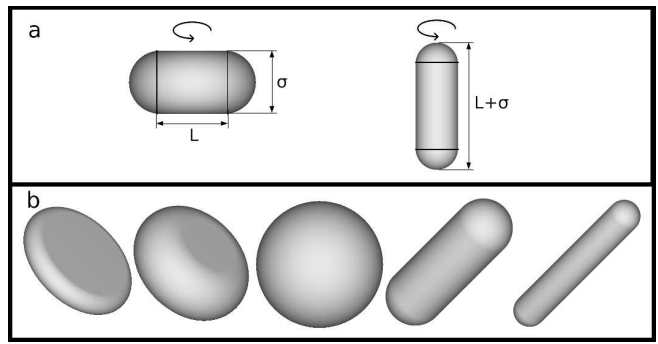}	
		\label{fig1}
	\caption{a) Schematic representation of an oblate (left) and prolate (right) solid of revolution. b) Examples of spherocylinders with aspect ratio (from left to right) equal to 0.1, 0.5, 1, 5, and 10.}
\end{figure}

To describe the inter-particle interactions, we have used the Soft Repulsive Spherocylinder (SRS) potential, which is obtained by truncating and shifting the Kihara potential \cite{CUE03} and was used in the past to model prolate \cite{CUE05, CUE15, CUE02}, and oblate mesogens \cite{MAR09B}. The SRS potential reads 

\begin{equation}\label{eq1}
U_{SRS} = \left\{ \begin{array}{cc}
4 \epsilon \left[  \left(\frac{\sigma}{d_{m}} \right)^{12} -
\left(\frac{\sigma}{d_{m}} \right)^6 + \frac{1}{4} \right] & ~~   \frac{\sigma}{d_{m}} \leq \sqrt[6]{2}\,
\\
0 & ~~ \frac{\sigma}{d_{m}} > \sqrt[6]{2}\ \end{array} \right.
\end{equation}

In the above equation, $\epsilon$ is the unit of energy, while $d_m$ is the minimum distance between the central cores of the particles, being a segment of length $L$ for prolate particles, and a disk of diameter $L$ for oblate particles. Efficient algorithms to calculate the minimum distance have been published for both particle geometries \cite{VEG90, CUE08}. As mentioned above, $\sigma$ represents the diameter of prolate particles as well as the thickness of oblate particles.

To study the diffusion of these model particles in nematic LCs, we have carried out Brownian dynamics (BD) simulations. The BD simulation technique is employed to mimic the behavior of suspensions of colloidal particles, whose size is significantly larger than that of the solvent molecules. Hence, the presence of the solvent is effectively incorporated by imposing a random drifting of the particles, whose trajectories are obtained by integrating the Langevin equation \cite{ALL87}. In BD simulations of non-spherical particles, the position, {\bf{r}}$_j$, and orientation, {\bf\^{u}}$_j$, of a particle $j$ over time $t$ are calculated by the following set of equations \cite{LOW94}: 

\begin{equation}
\label{eq2}
\begin{split}
{\bf r}^{\parallel}_j(t+\Delta t) =
{\bf r}^{\parallel}_j(t)+\frac{D^s_{\parallel}}{k_BT} {\bf F}^{\parallel}_j(t)\Delta t + \\
\,\,\,\,\,\,\,\,\,\, + (2D^s_{\parallel} \Delta t)^{1/2}
R^{\parallel} \textbf{\^{u}}(t)
\end{split}
\end{equation}

\begin{equation}
\label{eq3}
\begin{split}
{\bf r}^{\perp}_j(t+\Delta t) =    {\bf r}^{\perp}_j(t)+
\frac{D^s_{\perp}}{k_BT} {\bf F}^{\perp}_j(t)\Delta t+ \\
\,\,\,+ (2D^s_{\perp} \Delta t)^{1/2} (R^{\perp}_1 \textbf{\^{v}$_{j,1}$}(t)+ R^{\perp}_2 \textbf{\^{v}$_{j,2}$}(t)) \\
\end{split}
\end{equation}

\begin{equation}
\label{eq4}
\begin{split}
\textbf{\^{u}}_j(t+\Delta t) = \textbf{\^{u}}_j(t)+\frac{D^s_{\vartheta}}{k_BT} {\bf T}(t)\times
\textbf{\^{u}}(t)\Delta t+ \\
\,\,\,+ (2D^s_{\vartheta} \Delta t)^{1/2} (R^{\vartheta}_1
\textbf{\^{v}$_{j,1}$}(t)+R^{\vartheta}_2
\textbf{\^{v}$_{j,2}$}(t))
\end{split}
\end{equation}

\noindent where ${\bf r}^{\parallel}_j$ and ${\bf r}^{\perp}_j$ are the projections of the position vector ${\bf r}_j$ on the direction parallel and perpendicular to {\bf\^{u}}$_j$, respectively; ${\bf T}_j$ is the total torque acting over particle $j$ \cite{VEG90}; ${\bf F}^{\parallel}_j$ and ${\bf F}^{\perp}_j$ are the components of the force parallel and perpendicular, respectively, to {\bf\^{u}}$_j$; $R^{\parallel}$, $R^{\perp}_1$, $R^{\perp}_2$, $R^{\vartheta}_1$ and $R^{\vartheta}_2$ are independent Gaussian random numbers of variance 1 and zero mean; $\textbf{\^{v}}_{j,1}$ and ${\textbf{\^{v}}}_{j,2}$  are two random perpendicular unit vectors, being also perpendicular to vector $\textbf{\^{u}}_j$. The short time diffusion coefficients, $D^s_{\parallel}$, $D^s_{\perp}$ and $D^s_{\vartheta}$, have been calculated for both prolate and oblate particles with the analytical expressions proposed by Shimizu for spheroids \cite{SHI62}:

\begin{equation}\label{eq5}
\begin{split}
D^s_{\perp}&=D_{0}\frac{(2a^{2}-3b^{2})S+2a}{16\pi(a^{2}-b^{2})}b,\\
D^s_{\parallel}&=D_{0}\frac{(2a^{2}-b^{2})S-2a}{8\pi(a^{2}-b^{2})}b,\\
D^s_{\vartheta}&=3D_{0}\frac{(2a^{2}-b^{2})S-2a}{16\pi(a^{4}-b^{4})}b,
\end{split}
\end{equation}

In the above equations, $D_{0}= k_BT/\mu \sigma $, where $k_B$ is the Boltzmann constant, $T$ the absolute temperature, and $\mu$ the viscosity of the medium. $S$ is a geometric parameter that for prolate particles reads\cite{SHI62}

\begin{equation}
\label{eq6}
\begin{split}
\mbox{with\,\,\,\,\,}
S=\frac{2}{(a^{2}-b^{2})^{1/2}}\log\frac{a+(a^{2}-b^{2})^{1/2}}{b},\\
(a=\,(L+\sigma)/2,\,\,\,b=\,\sigma/2)\,\,\,\,\,\,\,\,\,\,\,\,\,\,\,\,\,\,\,\,
\end{split}
\end{equation}

and for oblate particles reads

\begin{equation}
\label{eq7}
\begin{split}
\mbox{with\,\,\,\,\,}
S=\frac{2}{(b^{2}-a^{2})^{1/2}}\arctan\frac{(b^{2}-a^{2})^{1/2}}{a},\\
(a=\,\sigma/2,\,\,\,b=\,L/2)\,\,\,\,\,\,\,\,\,\,\,\,\,\,\,\,\,\,\,\,
\end{split}
\end{equation}

Our interest is to compare the relative ability of oblate and prolate spherocylinders to diffuse in a nematic LC phase. To this end, one obvious choice would be to impose the same particle aspect ratio for the two geometries. Nevertheless, this choice, which allows one to consistently assess the phase behaviour of particles of different architectures, would not reproduce the same conditions of mobility at very short timescales or in extremely dilute suspensions. Because our aim is to have an insight into the effect of anisotropy on the long-time diffusion in structured fluids, we also equate the isotropic infinite-dilution translational, $D^s=(2D^s_{\perp}+D^s_{\parallel})/3$, or rotational, $D^s_{\vartheta}$, diffusion coefficients of oblate and prolate particles. To make this comparison we have chosen the aspect ratios $a_p=27$ and $a_o=0.1$ for the case of translational equivalence, and $a_p= 15.6$ and $a_o=0.1$ for the case of rotational equivalence. To fully clarify the effect of this choice, we have benchmarked our result with the case of identical shape anisotropy, where $a_p=1/a_o=15.6$.

For each of these three possible scenarios, we have started from an initial configuration of $N$ perfectly parallel particles randomly distributed in a cubic box, with $N=1260 - 2232$ at the desired packing fraction $\eta=\rho v_m$, being $\rho$ the numeric density of particles and $v_m$ the volume of the particles \cite{BOU86}. To equilibrate the system, we have run BD simulations of about $t=2000\tau$ for prolate and $t=20000\tau$ for oblate particles, where $\tau=\sigma^2/D_0$ is the time unit. The time step was fixed to $\Delta t=10^{-4}\tau$ for prolate and $10^{-5}\tau$ for oblate particles, while the temperature is $T^*=k_BT/\epsilon=1.465$ for both geometries. At this temperature the phase behavior of soft spherocylinders resembles that of a fluid of hard spherocylinders \cite{CUE05, CUE15}. The equilibration of the system has been monitored by checking the evolution of total energy and nematic order parameters. After equilibration, an additional BD simulation was carried out to compute a number of dynamical observables: (\textit{i}) the mean square displacement (MSD), (\textit{ii}) the self part of the intermediate scattering function (s-ISF), and (\textit{iii}) the four point susceptibility $\chi_4(q, t)$. In particular, the MSD is computed as follows 

\begin{equation} \label{eq8}
\big \langle \Delta r^{2}(t) \big \rangle= \Bigg \langle \frac{1}{N} \sum_{j=1}^{N} (\textbf{r}_j(t)-\textbf{r}_j(0))^2 \Bigg \rangle,
\end{equation}

\noindent where the brackets $\langle...\rangle$ denote ensemble average. We have also computed the MSD parallel, $\langle \Delta r^{2}_{\parallel}(t)\rangle$, and perpendicular, $\langle \Delta r^{2}_{\perp}(t)\rangle$, to the nematic director \textbf{\^{n}}. This vector has been calculated diagonalizing a symmetric traceless tensor incorporating the orientation vectors of all the particles \cite{ALL93}. The MSDs at long time scales are used to estimate the long time diffusion coefficients as follows:

\begin{equation}\label{eq9}
\begin{split}
D_{\perp}&=\lim_{t \to \infty} \frac{ \langle \Delta r^{2}_{\perp}(t)\rangle}{4t},\\
D_{\parallel}&=\lim_{t \to \infty} \frac{ \langle \Delta r^{2}_{\parallel}(t)\rangle}{2t},\\
D&= \lim_{t \to \infty} \frac{ \langle \Delta r^{2}(t)\rangle}{6t},
\end{split}
\end{equation}

\noindent We stress that these diffusion coefficients are different from those calculated in Eq.\  \ref{eq5}, which only take into account the effect of the solvent, but disregard the interaction with other colloidal particles. The s-ISF gives a measure of the structural relaxation decay of density fluctuations and reads:

\begin{equation}\label{eq10}
F_{s}(\textbf{q},t) = \frac{1}{N} \Bigg \langle \sum_{j=1}^{N} \exp[i \textbf{q} \cdot ( \textbf{r}_{j}(t+t_{0}) - \textbf{r}_{j}(t_{0})) ] \Bigg \rangle,
\end{equation}

\noindent where $\textbf{q}$ is the wave vector calculated at relevant peaks of the static structure factor and $\textbf{r}_{j}(t)$ is the particle position at time $t$. 









To explore the occurrence of collective motion, we have computed the four-point susceptibility function, $\chi_4(q, t)$, which measures the fluctuations of the s-ISF and provides information on the size and time evolution of the transient clusters formed in the fluid \cite{BER07, ABE07, MAR11B}. This function determines the eventual occurrence of collective motion by mapping the dynamics in two different spatial domains at two different times, hence its four-point nature. It is calculated as:

\begin{equation}\label{eq12}
\chi_4(q, t)= N \big[ \big \langle f^2_s({\bf q},t) \big \rangle - F_s^2({\bf q,t}) \big ]
\end{equation}

\noindent where $f_s({\bf q},t) =1/N \sum_{j=1}^{N} \cos({\bf q}[ ( \textbf{r}_{j}(t+t_{0}) - \textbf{r}_{j}(t_{0}))]$ is the real part of the instantaneous value of the s-ISF. All these dynamical observables have been averaged over trajectories with at least 100 independent time origins.


\section{Results}

In this section, we discuss the general characteristics of the diffusion in nematic fluids of prolate and oblate particles with (\textit{i}) equal aspect ratio, (\textit{ii}) equal infinite-dilution translational diffusion coefficients, and (\textit{iii}) equal infinite-dilution rotational diffusion coefficients. In the following, we refer to these three case scenarios as C$_1$, C$_2$, and C$_3$, respectively. The infinite-dilution diffusion coefficients, obtained from Eq.\ 5, are the same as those calculated in concentrated suspensions at very short time scales, when the particles are still rattling around their original position and have not yet interacted with their nearest neighbors.

 In Fig.\ 2, we report the parallel and perpendicular components of the MSD for oblate and prolate particles with shape anisotropy $a_p=a_o^{-1}=15.6$ in nematic LCs with packing fraction $\eta=0.35$. At very short time scales, with $t/\tau<0.5$, prolate particles diffuse faster in the direction perpendicular to $\bf\hat{n}$ and $\langle \Delta r^{2}_{\perp}(t)\rangle > \langle \Delta r^{2}_{\parallel}(t)\rangle$, which is coherent with the short time (or infinite-dilution) diffusion coefficients calculated from Eq.\ 5. At intermediate times, the perpendicular mobility becomes slower than the parallel mobility and an inversion in the trend observed at shorter times is observed. In particular, $\langle \Delta r^{2}_{\perp}(t)\rangle < \langle \Delta r^{2}_{\parallel}(t)\rangle$ at $t/\tau > 1$.
 
\begin{figure}[h]
	\center
	\includegraphics[width=0.48\textwidth]{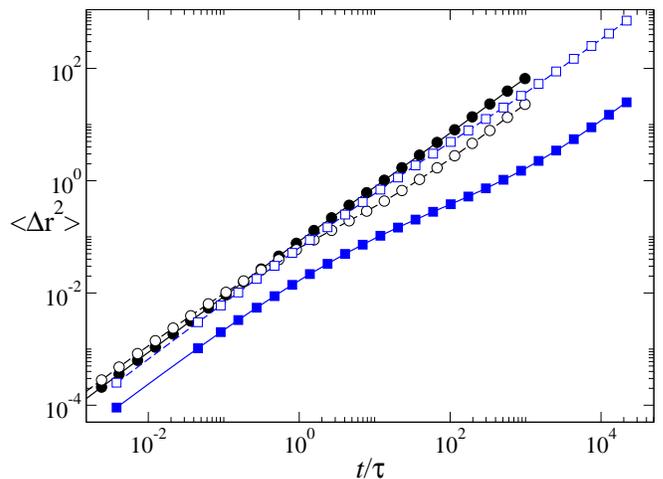}
	\label{fig2}	
	\caption{
		Parallel (solid lines and symbols) and perpendicular (dashed lines and open symbols) components of the MSD of prolate (black lines and circles) and oblate (blue lines and squares) particles with shape anisotropy $a_p=a_o^{-1}=15.6$ in nematic liquid crystals with packing fraction $\eta=0.35$.}
	\end{figure}

It is then possible to observe an intermediate regime where the slope of the  perpendicular component of the MSD decreases significantly. Although $\langle \Delta r^{2}_{\perp}(t)\rangle$ does not reach a clear plateau as previously observed in glasses \cite{BER14} or smectic phases \cite{PAT09, PAT10, MATENA}, the transport of prolate particles in this direction shows a sub-diffusive regime, indicating the time and length scales over which particle start to collide with other particles in the direction perpendicular to $\bf\hat{n}$. At long times, this component shows a diffusive regime, characterised by the long-time diffusion coefficient obtained from the slope of $\langle\Delta r^{2}_{\perp}(t)\rangle$. By contrast, the parallel component of the MSD shows a smoother behavior, with an almost insignificant variation of the slope at intermediate times. Consequently, at long times the diffusion of prolate particles is mainly in the direction parallel to the nematic director, being the main contribution to the total MSD (not shown here). In summary, the diffusion of prolate particles in nematic LCs is clearly anisotropic with a fast and slow component in the direction, respectively, parallel and perpendicular to $\bf\hat{n}$. The behavior of oblate particles with identical aspect ratio is characterised by opposite tendencies. Again, a clear anisotropy in the particle diffusion is observed. Nevertheless, the fast component is the one perpendicular to $\bf\hat{n}$, while the slow component, which exhibits a sub-diffusive regime at intermediate times, is parallel to it.

To analyse the relative ability of oblate and prolate particles of identical aspect ratio to diffuse in nematics, we calculate the total long-time diffusion coefficients, $D^+=(2D_{\perp}^+ + D_{\parallel}^+)/3$ and $D^-=(2D_{\perp}^- + D_{\parallel}^-)/3$, of prolate ($^+$) and oblate ($^-$) geometries, respectively. The resulting diffusivities show that the diffusion of prolate particles in the nematic phase is faster than that of oblate particles. This can be observed in Fig.\ 3, where we compare the parallel, perpendicular and total long-time diffusion coefficients in the nematic phase. For the sake of comparison, we also add the diffusivities obtained in the isotropic phase.

\begin{figure}[h]
	\center
	\label{fig3}	
	\includegraphics[width=0.48\textwidth]{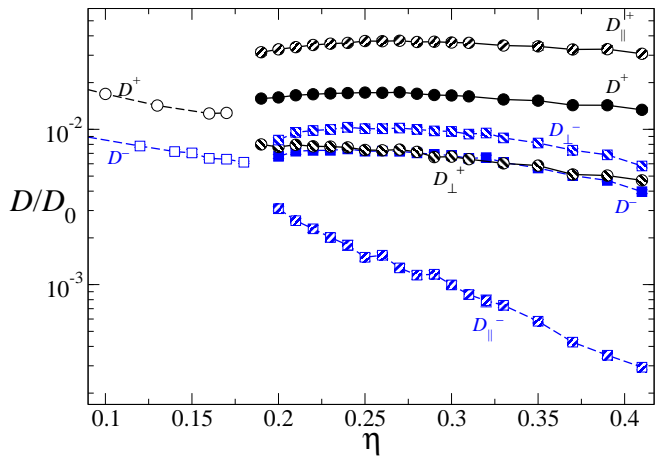}
	\caption{Parallel, perpendicular and total diffusion coefficients of prolate (circles) and oblate (squares) spherocylinders as a function of the packing fraction. Shape anisotropy: $a_p=a_o^{-1}=15.6$. Open symbols refer to the diffusion in the isotropic phase, whereas solid and striped symbols refer to the diffusion coefficients calculated in the nematic phase, as indicated by the labels.}
\end{figure}

As far as the nematic phase is concerned, the total diffusion coefficient of prolate particles is at least twice as large as that of oblate particles, depending on the packing of the phase. Of the same order of magnitude is the ratio between the main diffusivities of the two geometries, that is $D_{\parallel}^+ \approx 3 D_{\perp}^-$. If one compares the minor contributions to the total particle diffusivity, prolate geometries are still faster, with $D_{\perp}^+ > D_{\parallel}^-$, especially at particularly large packing fractions, where the diffusion of oblate spherocylinders along the director decreases dramatically. 
 
The dependence of the diffusivities in the nematic phase on the packing fraction does not appear to be monotonic. In particular, the total diffusion coefficients gradually increase with the system density up to a maximum, beyond which an inverse correlation is found. This tendency is the result of a non-monotonic behavior of the main contributions, $D_{\parallel}^+$ and $D_{\perp}^-$, to the total diffusivity, while the minor contributions, $D_{\perp}^+$ and $D_{\parallel}^-$, only decrease with increasing $\eta$. This behavior, specially clear in the case of oblate particles, had been reported before by de Miguel and co-workers for prolate particles with smaller anisotropy than that studied here \cite{DEM92}, and by Jabbari-Farouji for infinitely thin disks \cite{JAB12}. 

Between $\eta=0.15$ and 0.2, a phase transformation of the isotropic to the prolate nematic (N$^+$) or oblate nematic (N$^-$) phase is observed. The I-to-N phase transition produces an interesting increase of the diffusion coefficients, regardless the particle anisotropy. In Fig.\ 3, one can see that the diffusivity of prolate and oblate spherocylinders decreases in the isotropic phase with increasing packing fraction, but increases again when the phase transformation is produced. This abrupt change is most likely the consequence of the structural characteristics of the nematic phase. More specifically, the I-to-N$^+$ transition produces the formation of quasi unidimensional channels that act as preferential paths for particle diffusion. The effect is similar for oblate particles, although such preferential paths are found in planes perpendicular to $\bf \hat{n}$ and thus are quasi two-dimensional. By contrast, in the directions perpendicular to these channels, the probability of collisions between particles is significantly higher. Although such preferential channels boost the diffusion of prolate and oblate particles in the parallel and perpendicular direction, respectively, to $\bf\hat{n}$, at increasing packing fractions, they become narrower and narrower and end up hampering, rather than promoting, particle diffusion. This produces the reduction of the diffusion coefficients observed in Fig.\ 3. We notice that the maximum in $D$ is not observed in systems of infinitesimally thin disks \cite{JAB12}, where the preferential channels of the N$^-$ phase would never get thinner, at increasing packing fraction, than the particles themselves. 

In the light of these results, we now explore the diffusion of prolate and oblate spherocylinders whose short-time translational diffusion coefficients are the same, namely $D^{s,+}=D^{s,-}$ or, equivalently, $D^{s,+}_{\parallel}+2D^{s,+}_{\perp}=D^{s,-}_{\parallel}+2D^{s,-}_{\perp}$. This condition is satisfied by the prolate and oblate particle aspect ratios $a_p=27$ and $a_o=0.1$, respectively. Additionally, we have also investigated the case of oblate and prolate particles with identical short-time rotational diffusion coefficient, that is $D^{s,+}_{\vartheta}=D^{s,-}_{\vartheta}$. For the same oblate anisotropy, that is $a_o=0.1$, this condition is met for $a_p=15.6$. The common tendency observed in systems of oblate and prolate particles is that a more pronounced anisotropy (smaller $a_o$ or larger $a_p$, respectively) determines a slower dynamics along $\bf \hat{n}$ and perpendicularly to it. This behavior was expected. What we want to understand here is whether or not the relative mobility of prolate and oblate particles changes when an equivalence of diffusivities, rather than a geometric equivalence, is imposed. To this end, in Fig.\ 4, we report the ratio between the total MSD of prolate and oblate spherocylinders calculated for the three cases explored here. In particular, circles, triangles and squares show $R_\text{{MSD}}\equiv\langle \Delta r^2 \rangle_\text{p}/ \langle \Delta r^2 \rangle_\text{o}$ for the cases C$_1$, C$_2$, and C$_3$, respectively. The general trend unveils that $R_\text{{MSD}}$ decreases at intermediate times, more or less significantly for the three cases studied, and then increases again until a saturation plateau that is expected to be observed at long time scales, beyond our simulation time. Of particular relevance is the case scenario C$_2$, where $R_\text{{MSD}}<1$ over almost three time decades, specifically between $t/\tau=10^{-2}$ and $40$. In this time window, the diffusion of oblate particles is slightly faster than that of prolate particles, a behavior that is not observed in C$_1$ and C$_3$, where $R_\text{{MSD}}>1$ over the complete timeline explored. These two cases would indicate that prolate spherocylinders are constantly faster than oblate spherocylinders. However, this conclusion, as we show here, strongly depends on the assumptions made and does not hold if the short-time translational diffusivities of oblate and prolate spherocylinders are the same. 

\begin{figure}[h]
	\center
	\includegraphics[width=0.48\textwidth]{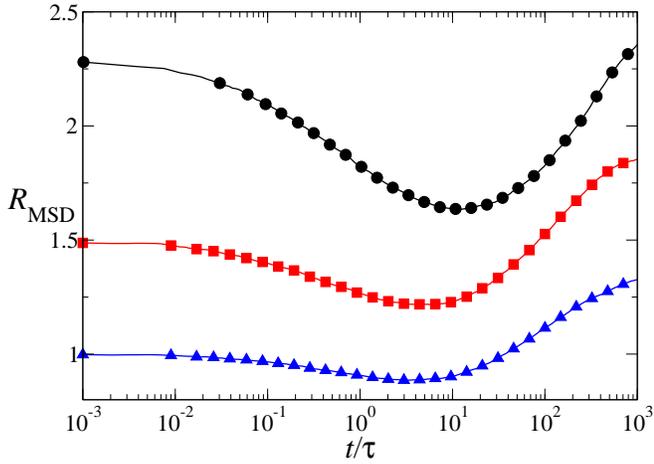}
	\label{fig4}
	\caption{Ratio between the MSD of prolate and oblate spherocylinders with identical short-time translational (blue triangles) and rotational (red squares) diffusion coefficients. For comparison, we also show $R_{\text{MSD}}$ for the case of identical aspect ratios (black circles). The packing fraction is $\eta= 0.35$ in all systems.}
\end{figure}





These considerations are confirmed by the analysis of the self-intermediate scattering function, $F_s(q, t)$, which quantifies the structural relaxation of the system over time. The s-ISFs of oblate and prolate spherocylinders for the cases C$_1$, C$_2$, and C$_3$ are displayed in Fig.\ 5, being all calculated at $|{\bf q}|=2\pi/\sigma$, corresponding to the position of the nearest neighboring particles. The complete set of s-ISFs shows a typical fluid-like behavior, with a single decay that is well fitted by a slightly stretched exponential function of the form $\exp[-(t/\tau)^\alpha]$, with $\alpha \approx 0.80$ and 0.88 for prolate and oblate spherocylinders, respectively.  Left and right frames, which refer to the cases C$_1$ and C$_3$, suggest a faster relaxation dynamics of prolate particles as compared to oblate particles. This is especially evident for C$_1$ (left frame), while for C$_3$ (right frame) the relaxation of both oblate and prolate particles is very similar, although slightly faster for the latter. An opposite trend is detected for C$_2$, where the $F_s$ decay of oblate particles slightly anticipates that of prolate particles. 

We have investigated the Gaussian dynamics of nematic LCs in a very recent work and refer the interested reader to Ref.\ \cite{CUE18} for further details. Here, we would like to stress that, due to anisotropic nature of nematic LCs, the above s-ISFs are expected to be approximated by an ellipsoidal, rather than spherical, Gaussian approximation, which incorporates the instantaneous values of the diffusion coefficients parallel and perpendicular to the nematic director \cite{CUE18}. This result was key to reconsider the supposed ubiquity of Fickian yet non-Gaussian dynamics, indeed detected in a wide spectrum of systems \cite{SHE05, WAN09, WAN12, KWO14, PHI15, THO16}, but not necessarily an intrinsic signature of soft matter.

\begin{figure}
	\center
	\includegraphics[width=0.48\textwidth]{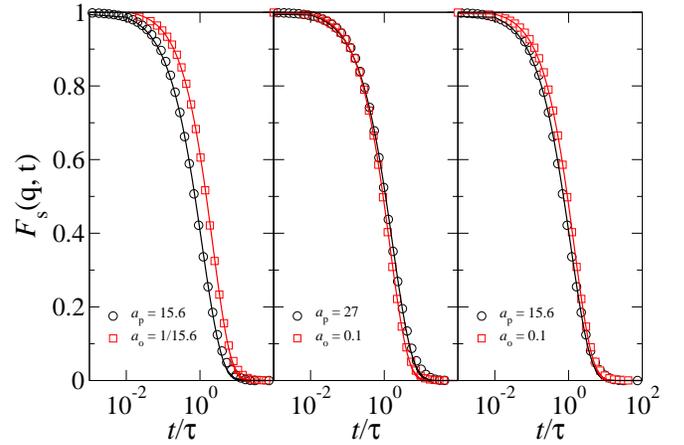}
	\caption{(Color online). Total self-intermediate scattering function $F_s(q, t)$ for prolate (black circles and lines) and oblate (red squares and lines) particles for the cases C$_1$ (left frame), C$_2$ (middle), and C$_3$ (right) in nematic liquid crystals with packing fraction $\eta= 0.35$. The solid lines are exponential fits.}
\end{figure}

 The stretched exponential decay of the self-ISFs would suggest a heterogenous signature of the long-time relaxation dynamics, with single particles trapped in transient cages formed by their neighbors. Two possible scenarios might explain such a non-exponential relaxation behavior: a heterogeneous scenario in which the particles relax exponentially at different relaxation rates, and a homogeneous scenario with the particles relaxing in a non-exponential manner at nearly identical rates \cite{Ediger, Richert}. In the latter case, a decreasing $\alpha$ would imply an increasing cooperativity, namely a collective motion of particles contributing to the relaxation of the system. 
 
 \begin{figure}
	\center
	\includegraphics[width=0.48\textwidth]{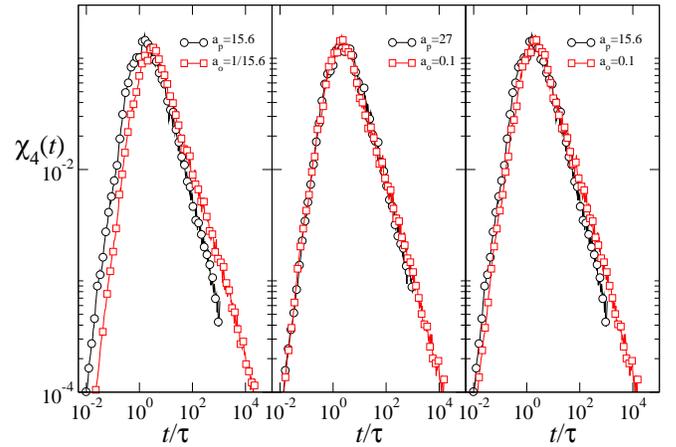}
	\caption{(Color online). 
		$\chi_4(q, t)$ at $q = 2\pi/\sigma$ for prolate (black circles and lines) and oblate (red squares and lines) particles for the cases C$_1$ (left frame), C$_2$ (middle), and C$_3$ (right) in nematic liquid crystals with packing fraction $\eta= 0.35$.}
\end{figure}
 
 Although the fitting coefficient $\alpha$ is not significantly lower than 1, we addressed the possible occurrence of a collective dynamical behavior by calculating the four-point susceptibility function, $\chi_4(q, t)$, which quantifies the magnitude of spontaneous fluctuations of the system dynamics as specified in Eq.\ (11) \cite{BER07, ABE07, MAR11B}. The resulting four-point susceptibilities calculated for oblate and prolate spherocylinders at $|{\bf q}|=2\pi/\sigma$, in C$_1$, C$_2$ and C$_3$ are shown in Fig.\ 6. Since $\chi_4(q, t)$ represents the average number of particles that are spatially correlated over time, its very small magnitude over the six decades explored  clearly indicates that the dynamics is not cooperative, regardless the  anisotropy and diffusion coefficients of the particles. In other words, the relaxation dynamics of nematic LCs relies entirely on the ability of individual particles to diffuse through their neighbors, with no sign a cooperative behavior as previously observed in smectic LCs \cite{PAT09, PAT10}.

\section{Conclusions}

In summary, we have investigated the dynamics in nematic liquid crystal phases of anisotropic particles, here modelled as oblate and prolate spherocylinders that confirm previous results \cite{JAB12,DEM92} and enrich the global picture of transport of particles in the nematic phase. In particular, our comparative study unveils that the generally accepted ability of prolate particles to diffuse faster than their oblate counterparts strongly depends on how this comparison is practically operated. In particular, we find that when prolate and oblate spherocylinders have identical infinite-dilution (or short-time) translational diffusion coefficients, the dynamics of oblate particles is faster in a significant time window as the MSD indicates. Additionally, under these conditions, the decay of the corresponding s-ISFs suggests a faster relaxation of systems of oblate particles. The diffusion in the nematic phase, regardless the particle geometry, displays a strong directional character, with a fast and slow component. It is remarkable that the relative relevance of the diffusion in the direction parallel and perpendicular to the nematic director is interchanged between prolate and oblate particles, being faster the diffusion parallel to the director in calamitic nematic fluids, while this role is taken by the perpendicular diffusion in discotic particles. The structural features of the nematic phase have a strong impact on the diffusion of particles, which abruptly increases across the isotropic-to-nematic transition. Despite the stretched exponential decay of the s-ISFs, which might imply the presence of a collective motion, the analysis of the four-point susceptibility function, $\chi_4(q,t)$, does not reveal any tangible signature of spatial correlations and thus excludes the occurrence of cooperative dynamics.

\begin{acknowledgements}	
AC and NM acknowledge project P12-FQM-2310 funded by the Junta de Andaluc\'{\i}a-FEDER and C3UPO for HPC facilities provided. AC also acknowledges grant PPI1719 awarded by the Pablo de Olavide University for funding his research visit to the School of Chemical Engineering and Analytical Science, The University of Manchester. AP acknowledges financial support from EPSRC under grant agreement EP/N02690X/1.
\end{acknowledgements}
\nocite{*}

\bibliography{neftalipatticuetos}

\end{document}